\documentclass[%
 aip,
 sd,%
 amsmath,amssymb,
 preprint,%
]{revtex4-1}

\usepackage{graphicx}
\usepackage{dcolumn}
\usepackage{bm}
\usepackage{color}

\begin{document}

\preprint{AIP/123-QED}

\title{Van der Waals epitaxial growth of topological insulator Bi$_{2-x}$Sb$_x$Te$_{3-y}$Se$_y$ ultrathin nanoplate on electrically insulating fluorophlogopite mica }

\author{Ngoc Han Tu}
\affiliation{Department of Physics, Graduate School of Science, Tohoku University 6-3, Aramaki Aoba-ku, Sendai, Miyagi, 980-8578 Japan}
 
\author{Yoichi Tanabe}%
 \email{youichi@sspns.phys.tohoku.ac.jp}
\affiliation{Department of Physics, Graduate School of Science, Tohoku University 6-3, Aramaki Aoba-ku, Sendai, Miyagi, 980-8578 Japan}

\author{Khuong Kim Huynh}
\affiliation{WPI-Advanced Institute for Materials Research, Tohoku University
2-1-1, Katahira Aoba-ku, Sendai, Miyagi, 980-0812 Japan}

\author{Yohei Sato}
\affiliation{Institute for Multidisciplinary Research for Advanced Materials, Tohoku University, Japan}

\author{Hidetoshi Oguro}
\affiliation{Institute for Multidisciplinary Research for Advanced Materials, Tohoku University, Japan}

\author{Satoshi Heguri}
\affiliation{WPI-Advanced Institute for Materials Research, Tohoku University
2-1-1, Katahira Aoba-ku, Sendai, Miyagi, 980-0812 Japan}

\author{Kenji Tsuda}
\affiliation{Institute for Multidisciplinary Research for Advanced Materials, Tohoku University, Japan}

\author{Masami Terauchi}
\affiliation{Institute for Multidisciplinary Research for Advanced Materials, Tohoku University, Japan}

\author{Kazuo Watanabe}
\affiliation{Institute for Multidisciplinary Research for Advanced Materials, Tohoku University, Japan}

\author{Katsumi Tanigaki}%
 \email{tanigaki@sspns.phys.tohoku.ac.jp}

\affiliation{Department of Physics, Graduate School of Science, Tohoku University 6-3, Aramaki Aoba-ku, Sendai, Miyagi, 980-8578 Japan}

\affiliation{WPI-Advanced Institute for Materials Research, Tohoku University2-1-1, Katahira Aoba-ku, Sendai, Miyagi, 980-0812 Japan}

\date{\today}

\begin{abstract}
We report the growth of high quality Bi$_{2-x}$Sb$_x$Te$_{3-y}$Se$_y$ ultrathin nanoplates (BSTS-NPs) on an electrically insulating fluorophlogopite mica substrate using a catalyst-free vapor solid method.
Under an optimized pressure and suitable Ar gas flow rate, we control the thickness, the size and the composition of BSTS-NPs.
Raman spectra showing systematic change indicate that the thicknesses and compositions of BSTS-NPs are indeed accurately controlled. 
Electrical transport demonstrates that a robust Dirac cone carrier transport in BSTS-NPs. 
Since BSTS-NPs provide superior dominant surface transport of the tunable Dirac cone surface states with negligible contribution of the conduction of the bulk states, BSTS-NPs provide an ideal platform to explore intrinsic physical phenomena as well as technological applications of 3-dimensional topological insulators in the future.

\end{abstract}

\maketitle

Three - dimensional topological insulators (3D TIs) are a new class of quantum matter in contemporary material physics \cite{1, 2}. 
The odd Z2 topological numbers protected by the specific crystal symmetry under strong spin-orbital coupling promises an existence of a gapless surface state, in which a Fermi circle encloses an odd number of Dirac points, leading to a quantized $\pi$ Berry's phase \cite{3}. 
This nontrivial topological phase in an electron wave function protects the surface metallic state from backward scattering and localization of itinerant electrons as long as the disorder potential does not violate the original electronic structure of the band inversion with preserving the time reversal symmetry \cite{4, 5}. 
While it was theoretically predicted that Bi-based chalcogenides (Bi$_2$Se$_3$, Bi$_2$Te$_3$) are prototypical TIs with perfect bulk insulating characteristics and a single Dirac cone on the surface \cite{6}, other carriers created by point defects and atomic displacement frequently tune their chemical potentials near the conduction or the valence band \cite{3}. 
Consequently, such bulk conduction channels often veil the intrinsic surface electric conduction of 3D-TIs.

An ultrathin film growth of 3D-TI materials has the potential to give an extremely idealistic condition for realizing dominant surface-transport relative to the bulk \cite{7, 8, 9, 10, 11}. 
In the case of ultrathin 3D-TI materials, the number of point defects can be significantly reduced \cite{12, 13}. 
Recently, Bi$_{2-x}$Sb$_x$Te$_{3-y}$Se$_y$ (BSTS) was demonstrated to be one of the most promising TIs by angle-resolved photoemission spectroscopy (ARPES) \cite{14} as well as electrical transport measurements \cite{15, 16}, and  satisfied the simultaneous requirement of both a highly insulating bulk and tunable Dirac cone carrier number on the surface.
Accordingly, the exquisite combination of an ultrathin film structure and an optimal chemical composition around the Dirac neutral point on the surface states in BSTS is highly essential to realize the ideal state of 3D-TIs from a material viewpoint.

Herein, we report on van der Waals epitaxial growth and the characterization of high quality BSTS ultrathin nanoplatelets (BSTS-NPs) on an electrically insulating fluorophlogopite mica (KMg$_3$(AlSi$_3$O$_{10}$)F$_2$) substrate having a pseudo hexagonal Z$_2$O$_5$ layered structure (Z = Si, Al). 
An atomically smooth surface of mica can be obtained by a cleavage along (001) plane, providing an ideal substrate for van der Waals epitaxy of layered topological insulators A$_2$B$_3$ (A = Bi, Sb; B = Se, Te) \cite{8}.
Using a special two-step growth condition for thin films, we can control not only the film thicknesses to a size as small as a few layers but also the compositions. 
Raman spectra showing a systematic change of peak positions, peak intensities and peak widths are the evidences that the thickness and composition of BSTS-NPs are indeed accurately controlled.
Electrical transport showing Shubnikov-de Haas (SdH) oscillations as the result of a finite Berry's phase demonstrates that a robust Dirac cone carrier transport in the vicinity of a Dirac neutral point is realized in BSTS-NPs. 
Since BSTS-NPs provide superior dominant surface transport of the tunable Dirac cone surface states with negligible contribution of the conduction of the bulk states, BSTS-NPs without any mechanical damage described in the present report will provide an ideal platform to explore intrinsic physical phenomena as well as technological applications of 3D TIs in the future.

BSTS-NPs were grown by a catalyst free vapor solid method, using a horizontal tube furnace equipped with a quartz tube \cite{7, 8}.
For growth of NPs, a Bi$_{1.5}$Sb$_{0.5}$Te$_{1.7}$Se$_{1.3}$ single crystal showing an insulating property reported in the previous report \cite{15} was employed as the source material.
The source material was placed in the upstream position of a furnace while a cleaved fluorophlogopite mica substrate was placed in the colder downstream position 11 cm away from the source material.
With a suitable Ar gas flow rate (60 - 200 sccm), the position of the source material was heated up to 500$^{\circ}$C and was kept at this temperature during the growth.
The position of the mica substrate was kept at 340 - 360 $^{\circ}$C.
In the present study, both pressure and flow rate of Ar gas were varied over the range of 10$^2$ - 10$^5$ Pa and 60 - 200 sccm during the growing process to realize efficient nucleation and growth of NPs on a fluorophlogopite mica substrate.
The composition of the BSTS-NPs was evaluated using energy dispersive x-ray spectroscopy (EDX) of JEOL JEM-2010 and JSM-7800F.
The thickness of the BSTS-NPs was determined using an atomic force microscope (AFM).
Raman spectra were measured using a micro Raman spectrometer (Labram HR-800, Horiba) with 628 nm laser excitation in the energy below 350 cm$^{-1}$ and cut-off energy around 70 cm$^{-1}$ by notch filters at room temperature.
The spectra were obtained through a 100~objective and recorded with 1800 lines/mm grating providing the spectral resolution of 1 cm$^{-1}$.
To investigate carrier transport in the Dirac cone surface states of the BSTS-NPs, electrical transport measurements were carried out using a semiconductor device analyzer (Agilent B1500) and a nano volt preamp for magnetic fields ($B$) below 17.5 T in the temperature range between 2- 300K.
In order to fabricate the electrodes using electron beam lithography, the BSTS-NPs were initially transferred from mica to a SiO$_2$/Si (SiO$_2$:300nm) substrate.

Accurate control in nucleation and growth of NPs was successively achieved by regulating  pressures and flow rates of Ar gas as shown in Figs 1 (a) - (e).
The nucleation of NPs  (step1: Fig. 1(f)) with sizes of around 8 $\mu$m locally occurred under atmospheric pressure (Fig. 1 (e)) with a high gas flow rate ($\sim$ 200 sccm), and subsequently a thin film grew at 10$^2$ Pa (Fig. 1 (b)).
The separation of nucleation from other subsequent growth processes was essential for obtaining single crystalline NPs.
During the growth process, the thickness was accurately controlled.
It is noted 20 -30 $\%$ variation of the thickness was observed along the scanned line of the AFM measurements with thicker part being closer to the edge of NP.
 In the step1, the size of NPs generally became large with an increase in thickness \cite{S}.
In order to control the thickness of NP, 10$^2$ Pa of Ar gas pressure was chosen for the growth condition of NPs (step2), since the two-step vapor phase crystal growth was found to be suitable for achieving high quality NPs.
In the step2, the nucleation continues on a substrate during an entire growth period under a high flow rate ($\sim$200 sccm), and the diffusion of the flux is larger than the migration rate of adatoms at the growth site. 
As a result, NPs grew eventually to their larger sizes starting from different seeds, with highly controlled thickness during the growth process (Fig. 1 (h)) \cite{S}.
Under a low gas flow rate ($\sim$ 60 sccm), on the other hand, a dynamic balance between diffusion and migration of the evaporated molecules was maintained during the growth.
Due to the high morphological anisotropy of BSTS crystals \cite{7}, growth in the lateral direction was preferred.
Accordingly, the high anisotropy observed in the growth process of BSTS was effective for controlling the size and thickness of ultrathin NPs (Fig. 1 (i)) \cite{S}.
Neither impurities nor phase separation were detected by energy-dispersive X-ray spectroscopy (EDX) in the present experiments.
It is noted that the nominal composition in the feed was changed to Bi$_{1.5}$Sb$_{0.5}$Te$_{1.7}$Se$_{1.3}$ under the Ar gas flow rate of 60 sccm and BiSbTeSe$_2$ under the rate of 200 sccm.

Figures 2 (b) and (c) show the dependence of Raman spectra on thickness of Bi$_{1.5}$Sb$_{0.5}$Te$_{1.7}$Se$_{1.3}$ and BiSbTeSe$_2$.
Raman spectra showed a systematic change from ultrathin NPs to bulk single crystals.
From AFM measurements, NP becomes thicker in the area closer to the edge.
In the present study, since the size of NP is substantially larger than the excitation laser spot size ($\sim$ 1 $\mu$m), spectra measured at the corresponding thickness positions were averaged to estimate the dependence of Raman spectra on thickness.

The crystal structure of tetradymite A$_2$B$_3$ (A = Bi, Sb, B = Se, Te) compounds consists of  a layered stacking sequence of B(1) - A - B(2) - A - B(1). 
In A$_2$B$_3$ compounds such as Bi$_2$Se$_3$, Bi$_2$Te$_3$ and Sb$_2$Te$_3$, four Raman modes are theoretically allowed under the D$_5^{3d}$ point group symmetry \cite{17, 18, 19-1, 19-2, 19}.
More than four broad humps were confirmed in our experiments although three Raman active modes are expected for BSTS.
Although the three inequivalent crystallographic sites in a unit cell are randomly occupied by Bi, Sb, Te or Se, the B(2) site may preferentially be occupied by Se due to its large electron negativity \cite{14}.
While Raman is a local probe of vibrational structure and is sensitively influenced by defects and atomic positions,\cite{20, 21} we may still expect that the Raman active modes may be influenced by the distribution of Bi, Sb or Te and Se at the A and B(1) sites.
In the present study, a Raman spectrum was simulated for the BiSbTe$_2$Se composition by employing an extended quintuple cell model where the random substitution is assumed under the minimum cell size \cite{S}.
The calculated frequencies of the Raman active phonon modes were listed in table S2 \cite{S}.
According to the calculations six Raman active modes exist in the range from 70 to 200 cm$^{-1}$.
Due to the limit of the lowest frequency ($\sim$ 70 cm $^{-1}$) in the present spectrometer, the detected lowest peak was not used for analyses.
Employing the five Raman active modes (2 E and 3 A1) observed in the range from 100 to 200 cm$^{-1}$, the Raman spectra of the two stoichiometric compounds were reproduced using a Lorentzian line shape function as shown in Figs. 2 (d) and (e).
The dependences of the peak positions, peak intensities and full width at half maximum (FWHM) of Raman peaks on NPs film thickness for Bi$_{1.5}$Sb$_{0.5}$Te$_{1.7}$Se$_{1.3}$ and BiSbTeSe$_2$ are listed in the table S3 and S4 \cite{S}.

While the number of obtained peaks was not changed, Bi$_{1.5}$Sb$_{0.5}$Te$_{1.7}$Se$_{1.3}$ and BiSbTeSe$_2$ showed different peak positions.
No clear change in the peak positions was observed for the five Raman modes, being independent from ultrathin NPs to bulk single crystals.
The dependences of the peak intensities on thickness of NPs for Bi$_{1.5}$Sb$_{0.5}$Te$_{1.7}$Se$_{1.3}$ and BiSbTeSe$_2$ are displayed in Figs. 2 (f) and (g).
The peak intensity continuously increased as the thickness decreased from bulk to ultrathin NP, being consistent with the multiple reflection effect of the excitation lasers between ultrathin NPs and insulating substrates \cite{19, 19-2, 19-3}.
The FWHM of Raman peaks tended to increase as the thickness decreases from bulk to ultrathin NP.
This behavior may be due to the decrease in phonon lifetimes of intralayer modes caused by the enhanced electron-phonon coupling in ultrathin NPs \cite{19, 19-4}.
The present results demonstrate that the composition and the thickness of BSTS-NPs can be characterized without any mechanical damages using peak positions and intensities, as well as FWHM of their Raman spectra.

Figures 3 (c) and (d) show the temperature dependences of electrical resistance ($R_{xx}$) for Bi$_{1.5}$Sb$_{0.5}$Te$_{1.7}$Se$_{1.3}$ (thickness: 100 nm) and BiSbTeSe$_2$ (thickness 105 nm).
A metallic behavior was confirmed below 50 K for Bi$_{1.5}$Sb$_{0.5}$Te$_{1.7}$Se$_{1.3}$ and below 100 K for BiSbTeSe$_2$, respectively, while both are semiconducting at high temperatures.
Since the surface conduction channel becomes dominant in electrical transport at low temperatures due to the reduction of the electron-phonon scattering rates, the metallic behaviors observed at low temperatures indicate that the quantum states of the surface Dirac cones of BSTS-NPs are responsible for the electrical transport.
Figure 3 (e) shows the magnetic field ($B$) dependences of $R_{xx}$.
While the convex-like curvature dominant in $R_{xx}$ at low Bs may originate from the weak anti-localization phenomenon \cite{22}, the Shubnikov de Haas (SdH) oscillations gradually become marked at high $B$s. 
The oscillatory modulations of $R_{xx}$ ($\Delta$ $R_{xx}$) against the inverse of $B$ (1/$B$) were deduced by employing a polynomial fitting for $R_{xx}$ by considering the corrections of its background magnetoresistance as shown in Figs. 3 (f) and (i).
By employing the fast Fourier-transformation, the oscillation frequencies were evaluated as shown in Figs. 3 (g) and (j).
The oscillations in $R_{xx}$($B$) can be described by

\begin{eqnarray}
 \Delta R_{xx} &\sim& \displaystyle{cos[2\pi(F/B - \gamma)]} \label{1},
\end{eqnarray}

where F is the oscillation frequency and 2$\pi$$\gamma$ is the phase factor \cite{23}.
The case of $\gamma$ = 1/2 shows the usual parabolic energy dispersion, while $\gamma$ = 0 is the Dirac cone-type linear dispersion due to the effect of the $\pi$ Berry's phase \cite{24}.
In order to estimate $\gamma$ in each oscillation, we plotted a Landau Level (LL) fan diagram by treating the valley (the peak) in $\Delta$ $R_{xx}$ as an integer n (a half integer $n$ = + 1/2 ) as shown in Figs. 3 (h) and (k). 
Applying a linear fit : $n$ = ($F$/$B$) + $\alpha$, $\alpha$ = 0.25 for Bi$_{1.5}$Sb$_{0.5}$Te$_{1.7}$Se$_{1.3}$ and $\alpha$ = 0.5 for BiSbTeSe$_2$ were obtained.
It was reported for BSTS that a small deviation in the energy dispersion from that of an ideal Dirac cone can be observed and the $B$ dependence of a Zeeman effect could induce a deviation in $\alpha$ from its ideal value \cite{25}. 
The present results of $\alpha$ = 0.25 - 0.5 were consistent with the topological $\pi$ Berry's phase surface in SdH oscillations.
It is noted that $\alpha$ = 0 detected for the other oscillatory part originates from a trivial quantum well due to the effect of band bending \cite{16, 26, 27, 28, 29}.

The Raman spectra of BSTS were investigated using an extended quintuple cell model, where the random distributions between Bi and Sb as well as between Te and Se are represented as A and B(1) sites. 
Although the effect of their random substitutions is suggested by Raman spectra, the topological electronic states of the TI materials are not strongly influenced by such randomly distributed potential scatterers.
In fact, the SdH oscillations characterized by the finite Berry's phase mean that  the robust Dirac cone metallic conduction is protected against both the backward scattering and the localization of electrons induced by the random potentials caused by the atomic positions.
The experimental facts demonstrate that high quality ultrathin BSTS-NPs could become a platform to investigate the intrinsic properties of 3D-TIs thanks to the large surface contributions of the Dirac cones in electrical transport with controllable Fermi energies. \cite{14, 15, 30}.

We described the growth of high quality ultrathin BSTS-NPs on an electrically insulating fluorophlogopite mica substrate using a catalyst-free vapor solid method.
Under an optimized pressure and suitable Ar gas flow rate, we controlled the thickness, the size and the composition of BSTS ultrathin films.
Employing Raman spectroscopy, the phonon modes of BSTS-NPs were evaluated using an extended quintuple cell model, where the random distribution between Bi and Sb as well as between Te and Se are assumed to be the R$\bar{3}$m space group.
The dependences of the peak positions and the intensities, as well as the FWHM of Raman peaks on the thickness and the composition of NPs were investigated.
As a result, BSTS-NPs were characterized without incurring mechanical damage using Raman spectra from the peak positions, the peak intensities and the FWHM of Raman peaks. 
The SdH oscillations involving a finite Berry's phase were confirmed.
Since the random distributions of substituent atoms did not change the topological macroscopic electric-transport properties, Raman spectra and SdH oscillations with finite Berry's phase are the essential aspects of BSTS-NPs.
Since the tunable Dirac cone inside of the semiconducting gap was demonstrated in BSTS \cite{14, 15, 16}, BSTS-NPs could be a platform to study the intrinsic physical properties of 3D-TIs, such as the Klein tunneling \cite{31} and the topological p-n junction \cite{32}.

We are grateful for H. Shimotani for the useful comments. We also grateful for A. Chatterjee and I. Kamatani to support simulations of Raman spectra using CASTEP program (Accelrys) and for S. Ikeda and R. Kumashiro (Common Equipment Unit of Advanced Institute for Materials Research, Tohoku University) to support the energy dispersive x-ray spectroscopy, Raman spectroscopy, and electron beam lithography experiments. This work was partly supported by the Grant-in-Aid for Young Scientists (B) (23740251), the Joint Studies Program (2013) of the Institute for Molecular Science and World Premier International Research Center Initiative (WPI), MEXT.


\begin{figure}[t]
\includegraphics[width=1.0\linewidth]{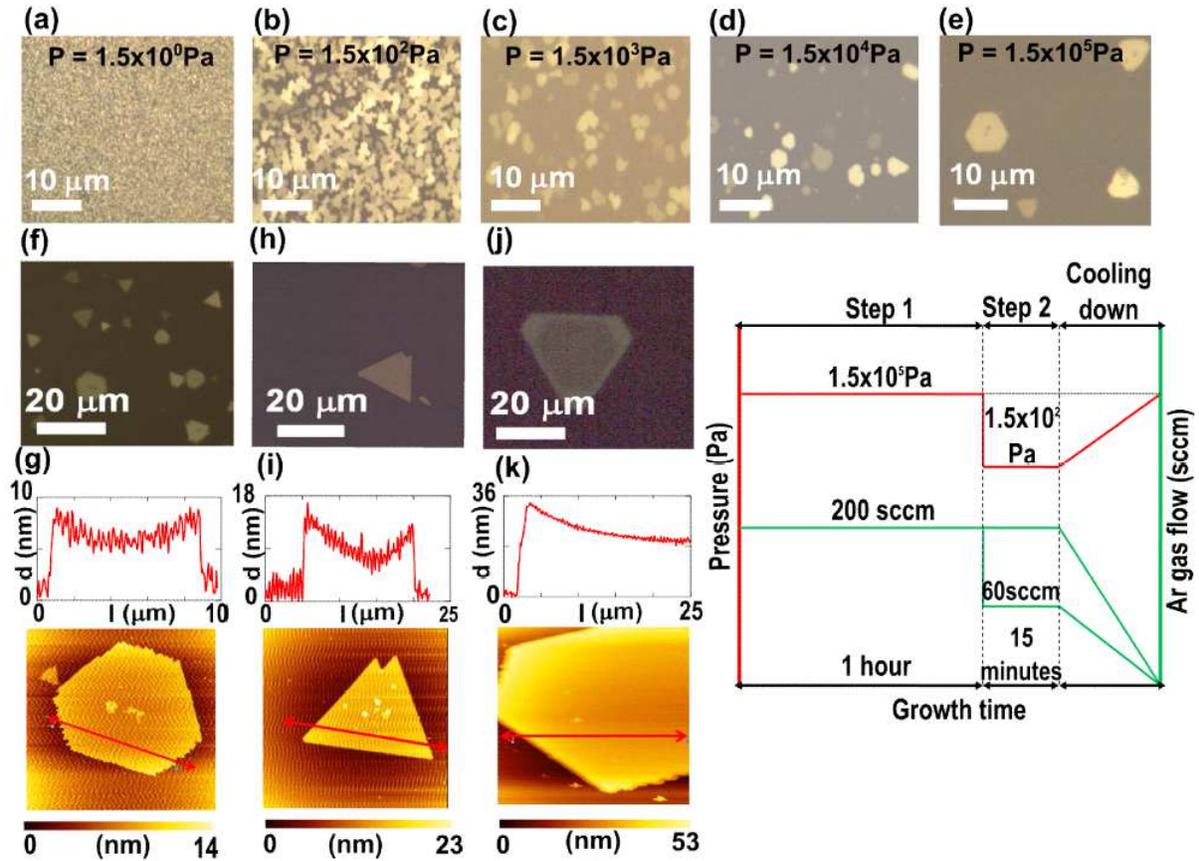}
\caption{
(a) - (e): Nucleation and growth of Bi$_{2-x}$Sb$_x$Te$_{3-y}$Se$_y$ (BSTS) ultrathin nanoplates (NPs) under various Ar gas pressures ($P$). 
The Ar gas flow rate ($F$) was fixed in 200 sccm. (f): $P$ = 1.5$\times$10$^5$  Pa, $F$ = 200 sccm (Step1), (h): $P$ = 1.5$\times$10$^5$  Pa , $F$ = 200 sccm (Step1) and $P$ = 1.5$\times$10$^2$  Pa, $F$ = 200 sccm (Step2), (j): $P$ = 1.5$\times$10$^5$  Pa, $F$ = 200 sccm, (Step1), $P$ = 1.5$\times$10$^2$ Pa, $F$ = 60 sccm, (Step2). (g), (i) and (k): Atomic force microscopy images of (f), (h), (j). (l): Growth diagram of BSTS-NPs.
}
\end{figure}

\begin{figure}[t]
\includegraphics[width=1.0\linewidth]{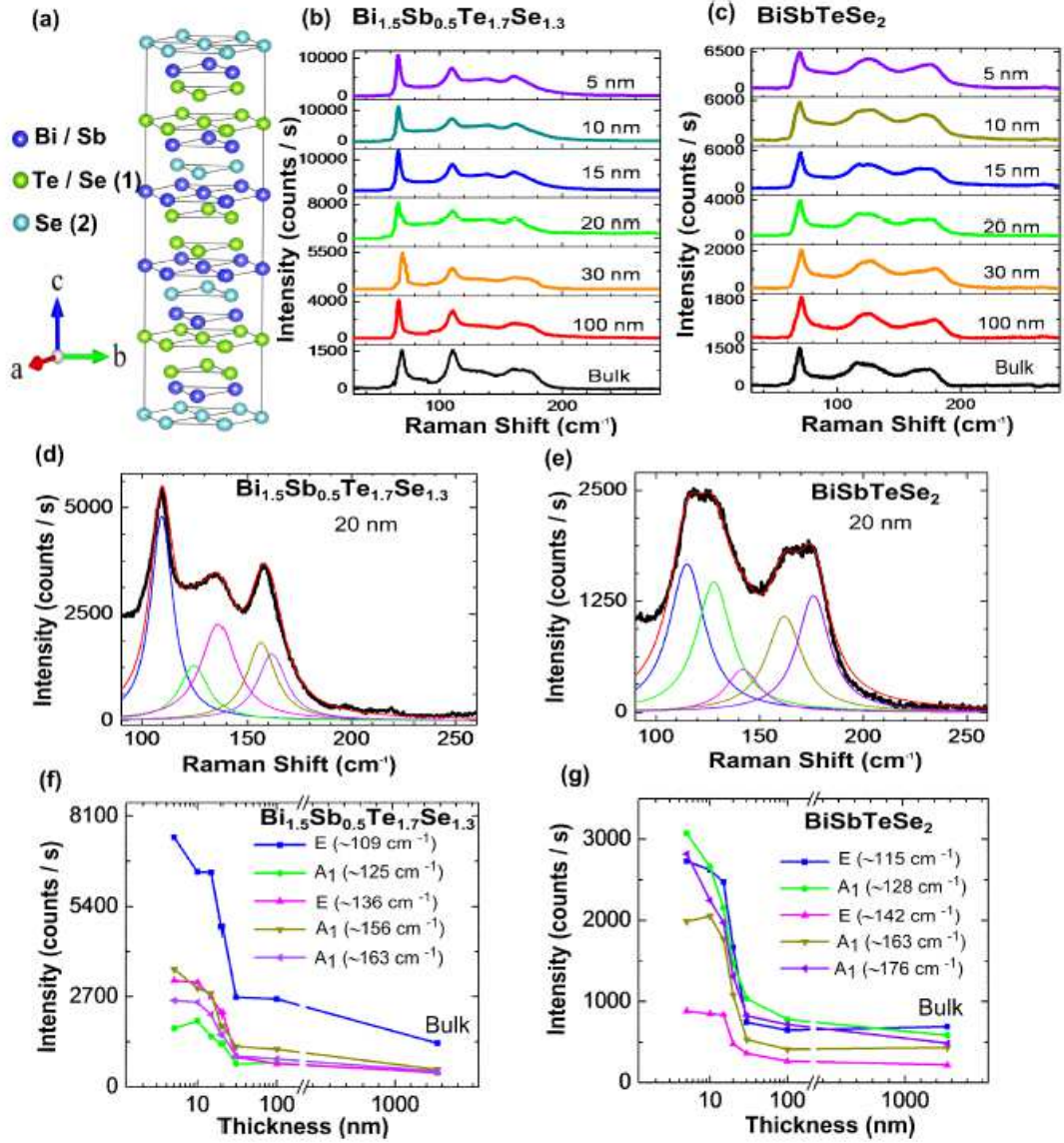}
\caption{(a) Crystal structure of Bi$_{2-x}$Sb$_x$Te$_{3-y}$Se$_y$ (BSTS).
Raman spectra of BSTS ultrathin nanoplates (NPs) with various thicknesses of (b) Bi$_{1.5}$Sb$_{0.5}$Te$_{1.7}$Se$_{1.3}$ (c) BiSbTeSe$_2$. 
Analytical results of Raman spectra of BSTS-NPs with a thickness of 20 nm (d) Bi$_{1.5}$Sb$_{0.5}$Te$_{1.7}$Se$_{1.3}$ (e) BiSbTeSe$_2$. 
Dependence of intensities on thickness of BSTS-NPs for the E and A1 modes of (f) Bi$_{1.5}$Sb$_{0.5}$Te$_{1.7}$Se$_{1.3}$ and  (g) BiSbTeSe$_2$.}
\end{figure}

\begin{figure}[t]
\includegraphics[width=1.0\linewidth]{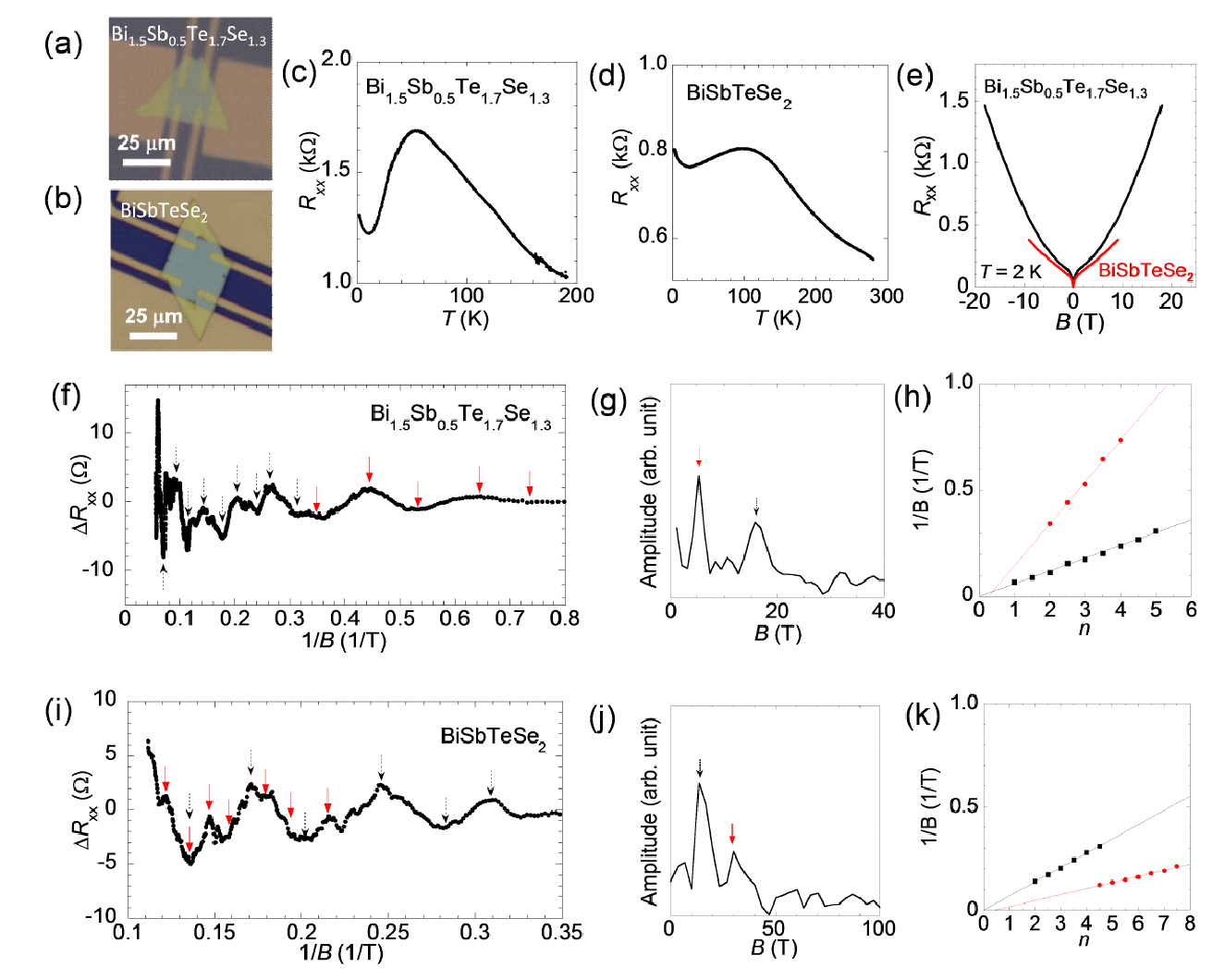}
\caption{Device pictures of (a) Bi$_{1.5}$Sb$_{0.5}$Te$_{1.7}$Se$_{1.3}$ and (b) BiSbTeSe$_2$ with thicknesses of 100 nm and 105 nm. Temperature dependence of electrical resistance ($R_{xx}$) for Bi$_{2-x}$Sb$_x$Te$_{3-y}$Se$_y$ (BSTS) ultrathin nanoplates (NPs): (a) Bi$_{1.5}$Sb$_{0.5}$Te$_{1.7}$Se$_{1.3}$ and (b) BiSbTeSe$_2$.
(c) Dependence of $R_{xx}$ on magnetic field ($B$).
Dependence of oscillatory part of $R_{xx}$ ($\Delta R_{xx}$) on 1/$B$ for BSTS-NP: (d) Bi$_{1.5}$Sb$_{0.5}$Te$_{1.7}$Se$_{1.3}$ and (e) BiSbTeSe$_2$.
Solid and broken arrows indicate the maximum and the minimum of the oscillations.
Results of fast Fourier transformation of 1/$B$ dependence of $\Delta$$R_{xx}$ for (f) Bi$_{1.5}$Sb$_{0.5}$Te$_{1.7}$Se$_{1.3}$ an (g) BiSbTeSe$_{2}$.
(h) and (i) Landau level Fan diagram plot employing a valley (a peak) in $\Delta$$R_{xx}$ as an integer $n$ (a half integer $n$+1/2 ).
Solid lines are results of linear fits.
}
\end{figure}

\begin{thebibliography}{32}\label{sec:TeXbooks}%
\bibitem{1}
M. Z. Hasan, C. L. Kane,
 Rev. Mod. Phys. {\bf 82}, 3045 (2010).

\bibitem{2}
X.- L. Qi, S.- C. Zhang,
 Rev. Mod. Phys. {\bf 83}, 1057 (2011).

\bibitem{3}
D. Hsieh, D. Qian, L. Wray, Y. Xia, Y. S. Hor, R. J. Cava, M. Z. Hasan,
 Nature 452, 970 (2008).

\bibitem{4}
S. Ryu, C. Mudry, H. Obuse, and A. Furusaki,
 Phys. Rev. Lett. {\bf 99}, 16601 (2007).

\bibitem{5}
K. Nomura, M. Koshino, S. Ryu,
 Phys. Rev. Lett. {\bf 99}, 146806 (2007).

\bibitem{6}
H. Zhang, C. -X. Liu, X. -L. Qi, X. Dai, Z. Fang, S. -C. Zhang,
 Nat. Phys. {\bf 5}, 438 (2009).

\bibitem{7}
D. Kong, W. Dang, J. J. Cha, H. Li, S. Meister, H. Peng, Z. Liu, Y. Cui,
 Nano Lett. {\bf 10}, 2245-2250 (2010).

\bibitem{8}
H. Li, J. Cao, W. Zheng, Y. Chen, D. Wu, W. Dang, K. Wang, H. Peng, Z. Liu,
 J. Am. Chem. Soc. {\bf 134}, 6132-6135 (2012).

\bibitem{9}
J. J. Cha, D. Kong, S.-S. Hong, J. G. Analytis, K. Lai, Y. Cui,
 Nano Lett. {\bf 12}, 1107-1111 (2012).

\bibitem{10}
P. Gehring, B. F. Gao, M. Burghard, K. Kern,
 Nano Lett., {\bf 12}, 5137-5142 (2012).

\bibitem{11}
D. Kong, K. J. Koski, J. J. Cha, S. S. Hong, Y. Cui,
 Nano Lett. {\bf 13}, 632-636 (2013).

\bibitem{12}
Y. Jiang, Y.Y. Sun, M. Chen, Y. Wang, Z. Li, C. Song, K. He, L. Wang, X. Chen, Q. K. Xue, X. Ma, S. B. Zhang,
 Phys. Rev. Lett., {\bf 108}, 066809 (2012).

\bibitem{13}
X. F. Kou, L. He, F. X. Xiu, M. R. Lang, Z. M. Liao, Y. Wang, A. V. Fedorov, X. X. Yu, J. S. Tang, G. Huang, X. W. Jiang, J. F. Zhu, J. Zou, K. L. Wang,
 Appl. Phys. Lett., {\bf 98}, 242102 (2011).

\bibitem{14}
T. Arakane, T. Sato, S. Souma, K. Kosaka, K. Nakayama, M. Komatsu, T. Takahashi, Z. Ren, K. Segawa, Y. Ando,
 Nat. Commun. {\bf 3}, 636 (2012).

\bibitem{15}
Z. Ren, A. A. Taskin, S. Sasaki, K. Segawa, Y. Ando,
 Phys. Rev. B {\bf 84}, 165311 (2011).

\bibitem{S}
See supplemental material at [].

\bibitem{16}
A. A. Taskin, Z. Ren, S. Sasaki, K. Segawa, Y. Ando,
 Phys. Rev. Lett. {\bf 107}, 016801 (2011).

\bibitem{17}
W.Richter and C. R. Becker,
 Phys.Stat.Sol(b), {\bf 84}, 619 (1977).

\bibitem{18}
S. L. Zhang, S. N. Wu, Y. Yan, T. Hu, J. Zhao, Y. Song, Q. Qu, W. Ding,
 J. Raman Specteosc. {\bf 39}, 1578-1583 (2008).

\bibitem{19-1}
D. Teweldebrahan, V. Goyal, A. A. Balandin, 
Nano Lett. {\bf 10}, 1209-1218 (2010).


\bibitem{19-2}
K. M. F. Shahil, M. Z. Hossain, D. Teweldebrhan, and A. A. Balandin, 
 Appl. Phys. Lett {\bf 96}, 153103 (2010).

\bibitem{19}
J. Zhang, Z.Peng, A. Soni, Y. Zhao, Y. Xiong, B. Peng, J. Wang, M. S. Dresselhaus, Q. Xiong,
 Nano Lett., {\bf 11}, 2407-2414 (2011).

\bibitem{20}
M. M. Lucchese, F. Stavale, E. H. Ferreira, C. Vilani, M. V. O. Moutinho, R. B. Capaz, C. A. Achete, A. Jorio,
 Carbon, {\bf 48}, 1592-1597 (2010).

\bibitem{21}
B. Guo, Q. Liu, E. Chen, H. Zhu, L. Fang, and J. R. Gong
 Nano Lett.{\bf 10},4975-4980 (2010).

\bibitem{19-3}
S. L. Li, H. Miyazaki, H. Song, H. Kuramochi, S. Nakaharai, and K. Tsukagoshi,
 ACS Nano {bf  6}, 7381-7388 (2012).

\bibitem{19-4}
R. He, Z. Wang, R. L. J. Qiu, C. Delaney, B. Beck, T. E. Kidd, C. C. Chancey, X. P. A. Gao, 
 Nanotechnology {\bf 23}, 455703 (2012).

\bibitem{22}
J. Chen, H. J. Qin, F. Yang, J. Liu, T. Guan, F. M. Qu, G. H. Zhang, J. R. Shi, X. C. Xie, C. L. Yang, K. H.Wu, Y. Q. Li,1, L. Lu,
 Phys. Rev. Lett., {\bf 105}, 176602 (2010).

\bibitem{23}
D. Shoenberg, Magnetic Oscillations in Metals (Cambridge University Press, Cambridge, 1984).

\bibitem{24}
G. P. Mikitik and Y.V. Sharlai,
 Phys. Rev. Lett., {\bf 82}, 2147 (1999).

\bibitem{25}
A. A. Taskin and Y. Ando,
 Phys. Rev. B, {\bf 84}, 035301 (2011).

\bibitem{26}
C. Z. Chang, K. He, L. L. Wang, X. C. Ma, M. H. Liu, Z. C. Zhang, X. Chen, Q. K. Xue,
 Spin, {\bf 1}, 21-25 (2011).

\bibitem{27}
G. Wang, X. G. Zhu, Y. Y. Sun, Y. Y. Li, T. Zhang, J. Wen, X. Chen, K. He, L. L. Wang, X. C. Ma, J. F. Jia, S. B. Zhang, Q. K. Xue,
 Adv. Mater. {\bf 2}, 2929-2932 (2011).

\bibitem{28}
Y. Zhang, K. He, C. Z. Chang, C. L. Song, L. L. Wang, X. Chen, J. F. Jia, Z. Fang, X. Dai, W. Y. Shan, S. Q. Shen, Q. Niu, X.L. Qi, S. C. Zhang, X. C. Ma, Q. K. Xue,
 Nat. Phys. {\bf 6}, 584-588 (2010).

\bibitem{29}
M. Bianchi, D.  Guan, S. Bao, J. Mi, B. B. Iversen, B.; P. D. C. King, P. Hofmann,
 Nat. Commun. {\bf 1}, 128 (2010).

\bibitem{30}
Y. Tanabe, K. K. Huynh, R. Nouchi, S. Heguri, G. Mu, J. T. Xu, H. Shimotani, K. Tanigaki,
 J. Phys. Chem. C {\bf 113}, 3533-3538 (2014).

\bibitem{31}
M. I. Katsnelson, K. S. Novoselov, A. K. Geim,
  Nat. Phys. {\bf 2}, 620-625 (2006).

\bibitem{32}
J. Wang, X. Chen, B. F. Zhu, S. C. Zhang,
 Phys. Rev. B {\bf 85}, 235131 (2012).


\end{thebibliography}
\end{document}